\documentstyle[aps]{revtex}

\begin{document}
\draft
\title{A dynamical evolution model on the black hole horizon}
\author{Jian-Yang Zhu\thanks{%
Electronic address: zhujy@bnu.edu.cn}$^{a,b}$}
\address{$^a$Department of Physics, Beijing Normal University, Beijing 100875, China\\
$^b$CCAST (World Laboratory), Box 8730, Beijing 100080, China}
\maketitle

\begin{abstract}
This paper demonstrates a dynamical evolution model of the black hole (BH)
horizon. The result indicates that a kinetic area-cells model of the BH's
horizon can model the evolution of BH due to the Hawking radiation, and this
area-cell system can be considered as an interacting geometrical particle
system. Thus the evolution turns into a problem of statistical physics. In
the present work, this problem is treated in the framework of
non-equilibrium statistics. It is proposed that each area-cell possesses the
energy like a microscopic black hole, and has the gravitational interaction
with the other area-cells. We consider both a non-interaction ideal system,
and a system with small nearest-neighbor interactions, and obtain an
analytic expression of the expected value of the horizon area of a dynamical
BH. We find that, after a long enough evolution, a dynamical BH with the
Hawking radiation can be in equilibrium with a finite temperature radiation
field. However, we also find that, the system has a critical point, and when
the temperature of the radiation field surrounding the BH approaches the
critical temperature of the BH, a critical slowing down phenomenon occurs.
\end{abstract}


The discovery of the thermal radiation of black holes (the Hawking radiation%
\cite{Hawking}) is a significant breakthrough in our understanding of black
holes. However, after three decades, there is still disagreement on a number
of important issues. For example, it is commonly accepted that black holes
have finite temperature, which is proportional to the surface gravity of the
event horizon and inversely proportional to the mass, and {\it Black holes
can be in equilibrium with a finite-temperature radiation field.} However,
it is also known that the thermal radiation of black holes is a quantum
effect. As the thermal radiation goes on, the black hole is emitting
particles. As its mass decreases, the temperature rises and, as a result,
the black hole emits particles more quickly. It is thus foreseeable that the 
{\it black hole cannot keep thermal equilibrium, and it will finally
evaporate and reach zero mass}. Now we turn to review the formation of a
black hole from the classical point of view. A star in its late stage of
evolution may evolve into a black hole, due to the gravitational collapse.
This black hole quickly reaches a stationary state, which can be
characterized by three parameters: mass, charge and angular momentum. Even
if we consider that the black holes with nonzero angular momentum and charge
are in an active excited state, {\it the Schwarzschild black holes with zero
angular momentum and zero charge are in an inactive ground state, which is
the end-result of star evolution}. In light of the mentioned-above points,
questions naturally arise: Certainly a black hole with a certain temperature
can be taken as a thermodynamic system, but can it be taken as a system in
thermal equilibrium? Is equilibrium statistics applicable? Will thermal
radiation really lead to the total evaporation of a black hole? What is the
end-result of star evolution? They are still open questions. Before the real
theory of quantum gravity is completely established, can we find consistent
answers?

The present work, we aim at the establishment of a model of non-equilibrium
dynamical evolution, in the framework of statistical physics. The expected
area of the event horizon is calculated, and its long time evolution is
observed. These results are used to shed light on the mentioned-above
questions.

{\it The model. }In the past years, Loop Quantum Gravity (LQG) has become a
serious candidate for a non-perturbative quantum theory of gravity\cite{LQG}%
. Its most notable prediction is the quantization of geometry\cite
{RovelliAshtekar}. As an analogy, in LQG, the fabric of space is like a
weave of tiny threads, and each thread poking through a surface gives it a
little bit of area. The surface area of a black hole (BH), then, is
generated by all the threads puncturing it. The event horizon is flat except
at these punctures, where it can flex, and the microstates of the BH are
defined by the different ways the event horizon can flex in or out\cite{Baez}%
.

According to the results of LQG, the horizon surface is split into a series
of $N$ discrete cells (small surface elements at the Planck scale). Each
cell has a discrete spectrum of area 
\begin{equation}
A_i(j_i)=8\pi l_p^2\gamma \sqrt{j_i\left( j_i+1\right) },i=1,2,3,...,N;\
j_i=0,1/2,1,3/2,\cdots ,  \label{A(j)}
\end{equation}
where $l_p$ is the Planck length, $\gamma $ is the Immirzi parameter, and $j$
is the irreducible representation of the gauge group $SU\left( 2\right) $.
Due to the Hawking quantum evaporation, the geometrical area of these
microscopic cells evolves with time.

Now, we can investigate the dynamical evolution of the event horizon of a BH
from the quantum point of view. During the time\footnote{%
With respect to ''time'', we must point it out that, although it is true
that there exists a notion of time in the 3+1 formulation, it is certainly
untrue that the notion of time is unambiguous. However, in order to carry on
the discussion of the dynamic evolution of black hole horizon using
non-equilibrium statistical mechanics, here we still give a clear definition
of ''time'' as an assumption. The extent to which this assumption is
acceptable can be judged from the final result of the research.} interval of 
$t\sim t+dt$, the event horizon of a macroscopic BH can be considered as a
set of area-cells at the Planck scale. The area-cells, with half-integer
spins, $j=0,1/2,$ $1,$ $3/2,$ $\cdots $ (irreducible representation of the
SU(2) gauge group), are characterized by quantized geometrical area, and
thus are referred to as {\it quantum geometrical particles}\footnote{%
As far as I know, the idea of ''quantum geometrical particle'' was first
given by Major and Setter in Ref. \cite{Major}. What I especially would like
to emphasize here is that, except the quantized geometric area (labeled by
spin $j$), the other microscopic details of the geometric particles have
been averaged out, in a way technically similar to the mean-field
approximation used to treat the ferromagnetic systems in condensed matter
physics. This is an issue worthy of further discussion.}. For an individual
area-cell, it is a small quantum BH, and its energy has the same form as a 
{\it Microscopic Black Hole}. However, adjacent area-cells have
gravitational interaction, and many area-cells together form an interacting
system of geometrical particles with spin. Since there can be various
possible transitions among the area-cells in different states, the
collective transition of the system becomes the quantum evaporation of the
macroscopic BH. We use a heat bath at temperature $T$ to simulate the
radiation field surrounding the BH, and suppose that the geometrical
particle system is in contact with this heat bath, but it may be in a
non-equilibrium state.

A system consisting of these geometrical particles with area (or spin) as a
characteristic variable is very similar to a spin-lattice system in
condensed matter physics. The contact of the system with a heat bath may
reflect the situation of a BH in a radiation field. The Hawking evaporation
leads the characteristic parameters to evolve with time. In this case, the
evolution of the event horizon of a BH can be investigated by studying the
dynamical behavior of a geometrical particle system in non-equilibrium
states. A generalized Glauber-type dynamics with single spin transition
mechanism\cite{Glauber,zhu1,zhu2} can be very useful here. In this paper, we
will reformulate the Glauber dynamics to make it suitable for the kinetic
geometrical system, and then apply it to a Black Hole with discrete area
spectrum.

{\it The formulation of the dynamical scheme. }The area of each fixed cells
on the event horizon of a BH is represented as a stochastic function of time 
$A_i(t)$, ($i=1,\cdot \cdot \cdot ,N$), which can be taken as discrete
values. Transitions can occur among these values. The transition probability 
$W_i(A_i(t)\rightarrow \hat{A}_i(t))$ from configuration $\left(
A_1(t),A_2(t),\cdots ,A_i(t),\cdots ,A_N(t)\right) $ to configuration $%
\left( A_1(t),A_2(t),\cdots ,\hat{A}_i(t),\cdots ,A_N(t)\right) $, in
general, depends on the momentary values of the neighboring cells as well as
on the influence of the heat bath. For this reason statistical correlations
exist between different geometrical particles. Therefore, it is necessary to
treat the entire $N$-particle system as a whole. The evolution of the
particles' area functions, which describes the evolution of the system,
forms a Markov process of $N$ discrete random variables with a continuous
time variable as argument.

We introduce a probability distribution function $P(\{A\};t)$, which denotes
the probability of the geometrical particle system being in the state of $%
\{A\}=(A_1,\cdots ,A_i,\cdots ,A_N)$ at time $t$. Let $W_i(A_i\rightarrow 
\hat{A}_i)$ be the transition probability per unit time that the $i$th
particle transits from area $A_i$ to another possible area $\hat{A}_i$,
while the others remain unchanged. Then, on the supposition of
single-particle transition, we may write the time derivative of the function 
$P(\{A\};t)$ as 
\begin{equation}
\frac d{dt}P(\{A\};t)=\sum_i\sum_{\hat{A}_i}g\left( \hat{A}_i\right) \left[
-W_i(A_i\rightarrow \hat{A}_i)P(\{A\};t)+W_i(\hat{A}_i\rightarrow
A_i)P(\cdots ,\hat{A}_i,\cdots ;t)\right] ,  \label{MasterEQ}
\end{equation}
where $g\left( \hat{A}_i\right) $ is the degeneracy of state $\hat{A}_i$.
This is a probability equation, in which the first term in the right-hand
side denotes the decrease of the probability distribution function $%
P(\{A\},t)$ per unit time due to the transition of the particle state from
the initial value $A_i$ $(i=1,2,\cdots ,N)$ to various possible final values 
$\hat{A}_i$, and the second term denotes the contrary situation. We shall
refer to Eq. (\ref{MasterEQ}) as the master equation since its solution
would contain the most complete description available of the system.

It is the most crucial step, obviously, to determine the transition
probabilities, $W_i(A_i\rightarrow \hat{A}_i)$, $i=1,2,3,\cdots ,N$, before
the master equation can be possibly solved. Then, how to determine the
transition probabilities? We have both mathematical and physical
considerations. On the one hand, mathematically, the transition probability
must be positive definite and normalized; and physically, a thermodynamic
system in a slowly varying process must have ergodicity and satisfy the
detailed balance condition. On the other hand, the transition probabilities
of the individual particles depend mainly on the momentary values of the
neighboring particles as well as on the influence of the heat bath. Thus,
the transition probability from $A_i$ to $\hat{A}_i$ must depend on the heat
Boltzmann factor of the neighboring particles. Based on these consideration,
the following form of the transition probability 
\begin{equation}
W_i(A_i\rightarrow \hat{A}_i)=\frac 1{Q_i}\exp \left[ -\beta {\cal H}%
_{eff}\left( \hat{A}_i,\sum_wA_{i+w}\right) \right]  \label{W}
\end{equation}
is a natural choice, where $Q_i$ is the normalization factor determined by
the normalized condition $\sum_{\hat{A}_i}g\left( \hat{A}_i\right)
W_i(A_i\rightarrow \hat{A}_i)=1$ and the summation $\sum_w$ is taken over
the neighboring particles of $i$.

Usually, we are only interested in the expected value of a certain function
of the area 
\begin{equation}
\left\langle f\left( A_k\right) \right\rangle =\sum_{\{A\}}g\left(
A_1\right) \cdots g\left( A_N\right) f\left( A_k\right) P(\{A\},t).
\label{def.}
\end{equation}
According to the definitions (\ref{def.}) and the master equation (\ref
{MasterEQ}), the time-evolution equation of $\left\langle f\left( A_k\right)
\right\rangle $ can be derived ( see Appendix) 
\begin{equation}
\frac d{dt}\left\langle f\left( A_k\right) \right\rangle =-\left\langle
f\left( A_k\right) \right\rangle +\sum_{\{A\}}g\left( A_1\right) \cdots
g\left( A_N\right) \left( \sum_{\hat{A}_k}g\left( \hat{A}_k\right) f\left( 
\hat{A}_k\right) W_k(A_k\rightarrow \hat{A}_k)\right) P(\{A\},t).
\label{evolving equation}
\end{equation}

{\it Application to the ideal geometrical particle system. }As a first step,
we consider an ideal geometrical particle system. In this case, the total
area of horizon, $A(t)=\sum_kA_k(t)$, can be taken as the parameter.
Therefore the time-evolution equation (\ref{evolving equation}) can be
simplified to the following form, 
\begin{equation}
\frac d{dt}\left\langle A(t)\right\rangle =-\left\langle A(t)\right\rangle
+\sum_Ag\left( A\right) \left( \sum_{\hat{A}}g\left( \hat{A}\right) \hat{A}%
W(A\rightarrow \hat{A})\right) P(A;t).
\label{ideal geometrical particle system}
\end{equation}
Because 
\begin{equation}
\sum_{\hat{A}}g\left( \hat{A}\right) \hat{A}W(A\rightarrow \hat{A})=\frac{%
\sum_{\hat{A}}g\left( \hat{A}\right) \hat{A}\exp \left[ -\beta {\cal H}%
_{eff}\left( \hat{A}\right) \right] }{\sum_{\hat{A}}g\left( \hat{A}\right)
\exp \left[ -\beta {\cal H}_{eff}\left( \hat{A}\right) \right] }%
=\left\langle A\right\rangle _{eq},  \label{expected value}
\end{equation}
and 
\begin{equation}
\sum_Ag\left( A\right) P(A;t)=TrP(A;t)\equiv 1,
\end{equation}
we can obtain 
\begin{equation}
\left\langle A(t)\right\rangle =\left\langle A\right\rangle _{eq}+\left[
\left\langle A(0)\right\rangle -\left\langle A\right\rangle _{eq}\right]
e^{-t}.  \label{solution}
\end{equation}

The expression (\ref{solution}) means that the event horizon as a whole
evolves in the form of exponential decrease with time. Obviously, the result
of a long-time evolution is $\left\langle A\left( \infty \right)
\right\rangle =\left\langle A\right\rangle _{eq}$. From Eq. (\ref{expected
value}), as long as the discrete spectrum expression of $A$ is given, the
expected value of the horizon's area, $\left\langle A\right\rangle _{eq}$,
can be calculated.

{\it Application to interacting geometrical particle system. }In our
consideration, an individual area-cell (a quantum geometrical particle) is
regarded as a microscopic quantum BH, and between neighboring geometrical
particles there is gravitational interaction. We can adopt as a hypothesis
that the energy (mass)-area relation is a power law, and as a specific
choice, we suppose that its energy (mass) is proportional to the square root
of the area (Schwarzschild-type)\footnote{%
What is chosen here is the same as Ref. \cite{Major} (see formula (5) of
this paper).}, 
\begin{equation}
\varepsilon _i=m_i\propto \sqrt{A_i(n_i)},
\end{equation}
where $A_i(n_i)=4\pi l_p^2\gamma \sqrt{n_i\left( n_i+2\right) }=a_0\sqrt{%
n_i\left( n_i+2\right) }$,$\ a_0=4\pi l_p^2\gamma $,$\
n_i=2j_i=0,1,2,3,\cdots $. The gravitational interaction between neighboring
geometrical particle, however, is yet an unknown problem. But, our
understanding is that: (1) the Planck scale $l_p$ is the minimum length in
the microscopic structure models of the quantum gravity theory; (2) the
event horizon is flat except at those punctures generated by all the spin
network's edges puncturing it\cite{Baez}. So, we might as well adopt simply
a classical gravitational potential as follows 
\begin{equation}
u\left( r_{ij}\right) =\left\{ 
\begin{array}{ll}
\infty \text{,} & \text{ }r_{i,j}=0, \\ 
-\frac{m_im_j}{r_{i,j}}\propto -\frac 1{l_p}\sqrt{A_i(n_i)}\sqrt{A_j(n_j)}%
\text{,} & \text{ }r_{i,j}=\text{NN distance,} \\ 
0\text{,} & \text{ otherwise}
\end{array}
\right.  \label{u}
\end{equation}
Thus, the effective Hamiltonian (with only nearest-neighbouring (NN)
interaction) can be written as 
\begin{equation}
{\cal H}_{\partial \Sigma }=\sum_i\varepsilon _i+\sum_{i<j}u\left(
r_{ij}\right) =m\sum_i\left[ n_i\left( n_i+2\right) \right]
^{1/4}-u\sum_{\left\langle i,j\right\rangle }\left[ n_i\left( n_i+2\right)
n_j\left( n_j+2\right) \right] ^{1/4},
\end{equation}
where $m$ and $u$ are the scaling factors, and $\ m>0$,$\ u>0$, and each sum 
$\left\langle i,j\right\rangle $ runs all NN pair. Thus, the transition
probability per unit time, that the characteristic quantity of the $i$th
particle transits from one value $n_i$ to another possible value $\hat{n}_i$%
, $n_i\rightarrow \hat{n}_i$, can be written as 
\begin{equation}
W_i\left( A_i\left( n_i\right) \rightarrow \hat{A}_i\left( \hat{n}_i\right)
\right) =\frac 1{Q_i}\exp \left\{ -\beta m\left[ \hat{n}_i\left( \hat{n}%
_i+2\right) \right] ^{1/4}+\beta u\sum_w\left[ \hat{n}_i\left( \hat{n}%
_i+2\right) n_{i+w}\left( n_{i+w}+2\right) \right] ^{1/4}\right\} ,
\end{equation}
where 
\[
Q_i=\sum_{\hat{n}_i}\left( \hat{n}_i+1\right) \exp \exp \left\{ -\beta
m\left[ \hat{n}_i\left( \hat{n}_i+2\right) \right] ^{1/4}+\beta
u\sum_w\left[ \hat{n}_i\left( \hat{n}_i+2\right) n_{i+w}\left(
n_{i+w}+2\right) \right] ^{1/4}\right\} . 
\]
Here, the degeneracy of spin state $j_i$, $g\left( \hat{\jmath}_i\right) =2%
\hat{\jmath}_i+1=\hat{n}_i+1$, is taken into account. At high temperature,
the summation for $\hat{n}_i$ can be approximately replaced by an integral $%
\sum_{\hat{n}_i}\rightarrow \int d\hat{n}_i$.

In order to get an analytical result, we consider the case of a weak
gravitational interaction and a high environmental temperature. Then, one
can obtain 
\begin{equation}
\sum_{\hat{n}_i}g\left( \hat{n}_i\right) \hat{A}_i\left( \hat{n}_i\right)
W_i(n_i\rightarrow \hat{n}_i)=\frac{20}{\beta ^2m^2}a_0\left[ 1+2\frac um%
\frac 1{\sqrt{a_0}}\sum_w\sqrt{A_{i+w}(n_{i+w})}+O\left( \left( \frac um%
\right) ^2\right) \right] ,
\end{equation}
\begin{equation}
\sum_{\hat{n}_i}g\left( \hat{n}_i\right) \sqrt{\hat{A}_i\left( \hat{n}%
_i\right) }W_i(n_i\rightarrow \hat{n}_i)=\frac 4{\beta m}\sqrt{a_0}\left[ 1+%
\frac um\frac 1{\sqrt{a_0}}\sum_w\sqrt{A_{i+w}(n_{i+w})}+O\left( \left( 
\frac um\right) ^2\right) \right] .  \label{21/7-1}
\end{equation}
Therefore, we have the following evolution equations 
\begin{equation}
\frac d{dt}\left\langle A_i(n_i)\right\rangle =20\frac{a_0}{\beta ^2m^2}%
-\left\langle A_i(n_i)\right\rangle +40\frac{\sqrt{a_0}}{\beta ^2m^2}\frac um%
\sum_w\left\langle \sqrt{A_{i+w}(n_{i+w})}\right\rangle +O\left( \left( 
\frac um\right) ^2\right) ,
\end{equation}
\begin{equation}
\frac d{dt}\left\langle \sqrt{A_i(n_i)}\right\rangle =4\frac{\sqrt{a_0}}{%
\beta m}-\left\langle \sqrt{A_i(n_i)}\right\rangle +4\frac 1{\beta m}\frac um%
\sum_w\left\langle \sqrt{A_{i+w}(n_{i+w})}\right\rangle +O\left( \left( 
\frac um\right) ^2\right) ;
\end{equation}
equivalently, 
\begin{eqnarray}
&&\frac d{dt}\left\{ \left\langle A(t)\right\rangle -\frac{10\sqrt{a_0}}{%
\beta m}\left[ \left\langle \sqrt{A_1(t)}\right\rangle +\left\langle \sqrt{%
A_2(t)}\right\rangle +\ldots \right] \right\}  \nonumber \\
&=&-\frac{20a_0}{\beta ^2m^2}N-\left\{ \left\langle A(t)\right\rangle -\frac{%
10\sqrt{a_0}}{\beta m}\left[ \left\langle \sqrt{A_1(t)}\right\rangle
+\left\langle \sqrt{A_2(t)}\right\rangle +\ldots \right] \right\} ,
\label{Eq-1}
\end{eqnarray}
\begin{eqnarray}
&&\frac d{dt}\left[ \left\langle \sqrt{A_1(t)}\right\rangle +\left\langle 
\sqrt{A_2(t)}\right\rangle +\ldots \right]  \nonumber \\
&=&\frac{4\sqrt{a_0}}{\beta m}N-\left( 1-\frac{16}{\beta m}\frac um\right)
\left[ \left\langle \sqrt{A_1(t)}\right\rangle +\left\langle \sqrt{A_2(t)}%
\right\rangle +\ldots \right] .  \label{Eq-2}
\end{eqnarray}
The exact solution is 
\begin{equation}
\left\langle A(t)\right\rangle =N\frac{20a_0}{\beta ^2m^2}\left( \frac 2{1-%
\frac{16}{\beta m}\frac um}-1\right) +\frac{10\sqrt{a_0}}{\beta m}%
C_2e^{-t/\tau }+C_1e^{-t},
\end{equation}
where $\left\langle A(t)\right\rangle =\sum_i\left\langle
A_i(t)\right\rangle $ denotes the total area, $C_1$ and $C_2$ are the
integral constants, and $\tau $ is the relaxation time of the system, 
\[
\tau =\frac 1{1-\frac{16}{\beta m}\frac um}=\left( 1-\frac{\beta _c}\beta
\right) ^{-1}=\left( 1-\frac T{T_c}\right) ^{-1}. 
\]
We notice that the system has a critical point, $\beta _c=\frac{16}m\frac um$%
. When the gravitational interaction is negligible, the system will evolve
rapidly to the equilibrium state, $\left\langle A\right\rangle _{eq}^0=N%
\frac{20a_0}{\beta ^2m^2}\propto l_p^2N\left( \frac{k_BT}m\right) ^2$, which
agrees with the result of Ref. \cite{Major}. However, a more complex
dynamical behavior will appear when the interaction cannot be ignored. If
the temperature of the radiation field is much lower than the critical
temperature $T_c$, the system will still rapidly approach the equilibrium
state $\left\langle A\right\rangle _{eq}=N\frac{20a_0}{\beta ^2m^2}\left( 
\frac 2{1-\frac{16}{\beta m}\frac um}-1\right) $. However, if the
temperature approaches the critical point, the system can hardly reach the
equilibrium state, and this phenomenon is commonly known as the critical
slowing down. It is surprising that the horizon area will increase
continuously when the temperature is higher than the critical point. But
this case should be excluded in the present discussion using this specific
method. The reason is that, due to the very intensive heat exchange, we
cannot regard the radiation field surrounding the BH as a heat bath with
constant temperature.

This paper demonstrates a dynamical evolution model of the black hole (BH)
horizon with discrete area spectrum. The result indicates that the evolution
of BH due to the Hawking radiation can be modeled by a kinetic area-cell
model of the BH's horizon, and this area-cell system can be considered as an
interacting geometrical particle system with spin. Thus the evolution turns
into a problem of statistical physics. In the present work, this problem is
treated in the framework of non-equilibrium statistics, and the expected
area of the event horizon is obtained. We find that, after a long enough
evolution, a dynamical BH with the Hawking radiation can be in equilibrium
with a finite temperature radiation field. However, we also find that, the
system has a critical point, and when the temperature of the radiation field
surrounding the BH approaches the critical temperature of the BH, a critical
slowing down phenomenon occurs.

Of course, the present work is only a preliminary attempt on the evolution
of the BH's horizon in the framework of a non-equilibrium statistics. Any
further study, such as to choose a well-defined Hamiltonian, and to consider
the possibility of creation or annihilation of geometrical particles, and so
on, is very interesting.

I close this paper with some comments. Due to the discreteness of spacetime
itself at the Planck scale, there is a minimum length, namely the Planck
length, which is similar to the lattice constant in condensed matter. Thus,
mature methods and viewpoints developed from condensed matter physics and
statistical physics can be used for reference in the study of the discrete
quantum spacetime. In this direction, some groups have made interesting and
enlightening efforts (see, for example, Refs. \cite
{Markopoulou-1,Markopoulou-2,Requardt-1,Requardt-2}).

\acknowledgments

The work was supported by the National Natural Science Foundation of China
(No. 10375008), and the National Basic Research Program of China
(2003CB716302).

\appendix 

\section{Proof of Eq. (5)}

\label{app-a}According to the definition (\ref{def.}) of $\left\langle
f\left( A_k\right) \right\rangle $ and the master equation (\ref{MasterEQ}),
we have 
\begin{eqnarray}
\frac d{dt}\left\langle f\left( A_k\right) \right\rangle &=&\sum_{\left\{
A\right\} }\left( \prod_{\alpha =1}^Ng\left( A_\alpha \right) \right)
f\left( A_k\right) \sum_i\sum_{\hat{A}_i}g\left( \hat{A}_i\right) \left[
-W_i\left( A_i\rightarrow \hat{A}_i\right) P\left( \left\{ A\right\}
,t\right) \right.  \nonumber \\
&&\left. +W_i\left( \hat{A}_i\rightarrow A_i\right) P\left( \left\{ A_{j\neq
k}\right\} ,\hat{A}_k,t\right) \right]  \nonumber \\
&=&\sum_{\left\{ A\right\} }\left( \prod_{\alpha =1}^Ng\left( A_\alpha
\right) \right) f\left( A_k\right) \sum_{i\left( i\neq k\right) }\sum_{\hat{A%
}_i}g\left( \hat{A}_i\right) \left[ -W_i\left( A_i\rightarrow \hat{A}%
_i\right) P\left( \left\{ A\right\} ,t\right) \right.  \nonumber \\
&&\left. +W_i\left( \hat{A}_i\rightarrow A_i\right) P\left( \left\{ A_{j\neq
k}\right\} ,\hat{A}_k,t\right) \right]  \nonumber \\
&&+\sum_{\left\{ A\right\} }\left( \prod_{\alpha =1}^Ng\left( A_\alpha
\right) \right) f\left( A_k\right) \sum_{\hat{A}_k}g\left( \hat{A}_k\right)
\left[ -W_k\left( A_k\rightarrow \hat{A}_k\right) P\left( \left\{ A\right\}
,t\right) \right.  \nonumber \\
&&\left. +W_k\left( \hat{A}_k\rightarrow A_k\right) P\left( \left\{ A_{j\neq
k}\right\} ,\hat{A}_k,t\right) \right] .  \label{A.1}
\end{eqnarray}
Looking at the $i\neq k$ term of (\ref{A.1}), 
\begin{eqnarray}
\mbox{$(i\neq k)$ term} &=&\sum_{\left\{ A_{j\neq i}\right\} }\left(
\prod_{\alpha =1(\alpha \neq i)}^Ng\left( A_\alpha \right) \right) f\left(
A_k\right)  \nonumber \\
&&\times \sum_{i\left( i\neq k\right) }\left[ -\sum_{A_i,\hat{A}_i}g\left(
A_i\right) g\left( \hat{A}_i\right) W_i\left( A_i\rightarrow \hat{A}%
_i\right) P\left( \left\{ A\right\} ,t\right) \right.  \nonumber \\
&&\left. +\sum_{A_i,\hat{A}_i}g\left( A_i\right) g\left( \hat{A}_i\right)
W_i\left( \hat{A}_i\rightarrow A_i\right) P\left( \left\{ A_{j\neq
k}\right\} ,\hat{A}_k,t\right) \right] ,  \label{A-2}
\end{eqnarray}
it is easy to see that this term equals to zero, as long as $\hat{A}_i$
exchange with $A_i$ before doing sum for $\hat{A}_i$ and $A_i$. So the
surplus term of (\ref{A.1}) is only the last term ($i=k$): 
\begin{eqnarray*}
\frac d{dt}\left\langle f\left( A_k\right) \right\rangle &=&\sum_{\left\{
A\right\} }\left( \prod_{\alpha =1}^Ng\left( A_\alpha \right) \right)
f\left( A_k\right) \sum_{\hat{A}_k}g\left( \hat{A}_k\right) \left[
-W_k\left( A_k\rightarrow \hat{A}_k\right) P\left( \left\{ A\right\}
,t\right) \right. \\
&&\left. +W_k\left( \hat{A}_k\rightarrow A_k\right) P\left( \left\{ A_{j\neq
k}\right\} ,\hat{A}_k,t\right) \right] \\
&=&-\sum_{\left\{ A\right\} }\left( \prod_{\alpha =1}^Ng\left( A_\alpha
\right) \right) f\left( A_k\right) \left( \sum_{\hat{A}_k}g\left( \hat{A}%
_k\right) W_k\left( A_k\rightarrow \hat{A}_k\right) \right) P\left( \left\{
A\right\} ,t\right) \\
&&+\sum_{A_1,\cdots ,A_k,\hat{A}_k\cdots ,A_N}g\left( A_1\right) \cdots
g\left( A_k\right) g\left( \hat{A}_k\right) \cdots g\left( A_N\right)
f\left( A_k\right) \\
&&\times W_k\left( \hat{A}_k\rightarrow A_k\right) P\left( \left\{ A_{j\neq
k}\right\} ,\hat{A}_k,t\right) \\
&=&-\left\langle f\left( A_k\right) \right\rangle +\sum_{\left\{ A\right\}
}\left( \prod_{\alpha =1}^Ng\left( A_\alpha \right) \right) \left( \sum_{%
\hat{A}_k}g\left( \hat{A}_k\right) f\left( \hat{A}_k\right) W_k\left(
A_k\rightarrow \hat{A}_k\right) \right) P\left( \left\{ A\right\} ,t\right)
\end{eqnarray*}
in which, the normalized condition $\sum_{\hat{A}_k}g\left( \hat{A}_k\right)
W_k\left( A_k\rightarrow \hat{A}_k\right) =1$ and the technique of exchange
of $\hat{A}_i$ for $A_i$ were used. Hitherto, Eq.(\ref{evolving equation})
have been proven exactly.

\end{document}